\documentclass[runningheads,envcountsame,oribibl]{llncs}
\usepackage[colorinlistoftodos,prependcaption,textsize=tiny]{todonotes}

\usepackage{amssymb}
\usepackage{amsfonts}
\usepackage{amsmath}
\usepackage{dsfont} 
\usepackage{graphicx}
\usepackage{lineno}
\usepackage{rotating}
\makeatletter
\newcommand{\figcaption}{\def\@captype{figure}\caption}
\newcommand{\tabcaption}{\def\@captype{table}\caption}
\makeatother

\usepackage[caption=false]{subfig}

\usepackage[misc]{ifsym}  
\usepackage{booktabs}
\usepackage{ctable}
\newcolumntype{+}{>{\global\let\currentrowstyle\relax}}
\newcolumntype{^}{>{\currentrowstyle}}

\newcommand{\PreserveBackslash}[1]{\let\temp=\\#1\let\\=\temp}
\newcolumntype{C}[1]{>{\PreserveBackslash\centering}p{#1}}
\newcolumntype{R}[1]{>{\PreserveBackslash\raggedleft}p{#1}}
\newcolumntype{L}[1]{>{\PreserveBackslash\raggedright}p{#1}}

\usepackage{makecell}
\usepackage{multirow}

\usepackage{microtype}
\usepackage[colorlinks=true,citecolor=blue,urlcolor=black]{hyperref}
\usepackage[capitalize]{cleveref}

\usepackage{soul}

\usepackage{mathtools}
\newtagform{myform}{\normalsize(}{)}
\usetagform{myform}

\usepackage{microtype}
\usepackage[colorlinks=true,citecolor=blue,urlcolor=black]{hyperref}

\usepackage{adjustbox}

\newcommand\blfootnote[1]{%
  \begingroup
  \renewcommand\thefootnote{}\footnote{#1}%
  \addtocounter{footnote}{-1}%
  \endgroup
}

\begin{document}

\title{Data consistency networks for (calibration-less) accelerated parallel MR image reconstruction}

\titlerunning{ISMRM Abstract \#4663: Data consistency networks for accelerated pMRI}  

\newcommand{\corrauth}{\textsuperscript{(\Letter)}}
\author{Jo Schlemper\inst{1}\corrauth \and Jinming Duan\inst{1} \and Cheng Ouyang\inst{1} \and Chen Qin \inst{1} \and Jose Caballero \inst{1} \and Joseph V. Hajnal\inst{2} \and Daniel Rueckert\inst{1}}
\authorrunning{J. Schlemper et al.}
\institute{Biomedical Image Analysis Group, Imperial College London, UK \email{\{jo.schlemper11,d.rueckert\}@imperial.ac.uk}
\and Imaging and Biomedical Engineering Clinical Academic Group, King's College London, UK\\ \email{\{jo.hajnal\}@kcl.ac.uk}
}

\mainmatter
\maketitle              

\begin{abstract}
We present simple reconstruction networks for multi-coil data by extending deep cascade of CNN’s and exploiting the data consistency layer. In particular, we propose two variants, where one is inspired by POCSENSE and the other is calibration-less. We show that the proposed approaches are competitive relative to the state-of-the-art both quantitatively and qualitatively.
\end{abstract}

\section{Introduction}

Recently, several deep learning approaches have been proposed for accelerated parallel MR image reconstruction\cite{hammernik2018learning,mardani2018deep,han2019k,cheng2018deepspirit,akccakaya2019scan,zhang2018multi}. In this work, we present simple reconstruction networks for multi-coil data by extending deep cascade of CNN’s\cite{schlemper2017deep}. In particular, we propose two approaches, where one is inspired by POCSENSE\cite{samsonov2004pocsense} and the other is calibration-less. The method is evaluated using a public knee dataset containing 100 subjects\cite{hammernik2018learning}. We show that the proposed approaches are competitive relative to the state of the art both quantitatively and qualitatively.
\blfootnote{Presented at ISMRM 27th Annual Meeting \& Exhibition (Abstract \#4663)}

\section{Methods}

\begin{figure}[t!]
	\centering
  \subfloat{
       \includegraphics[width=0.48\linewidth]{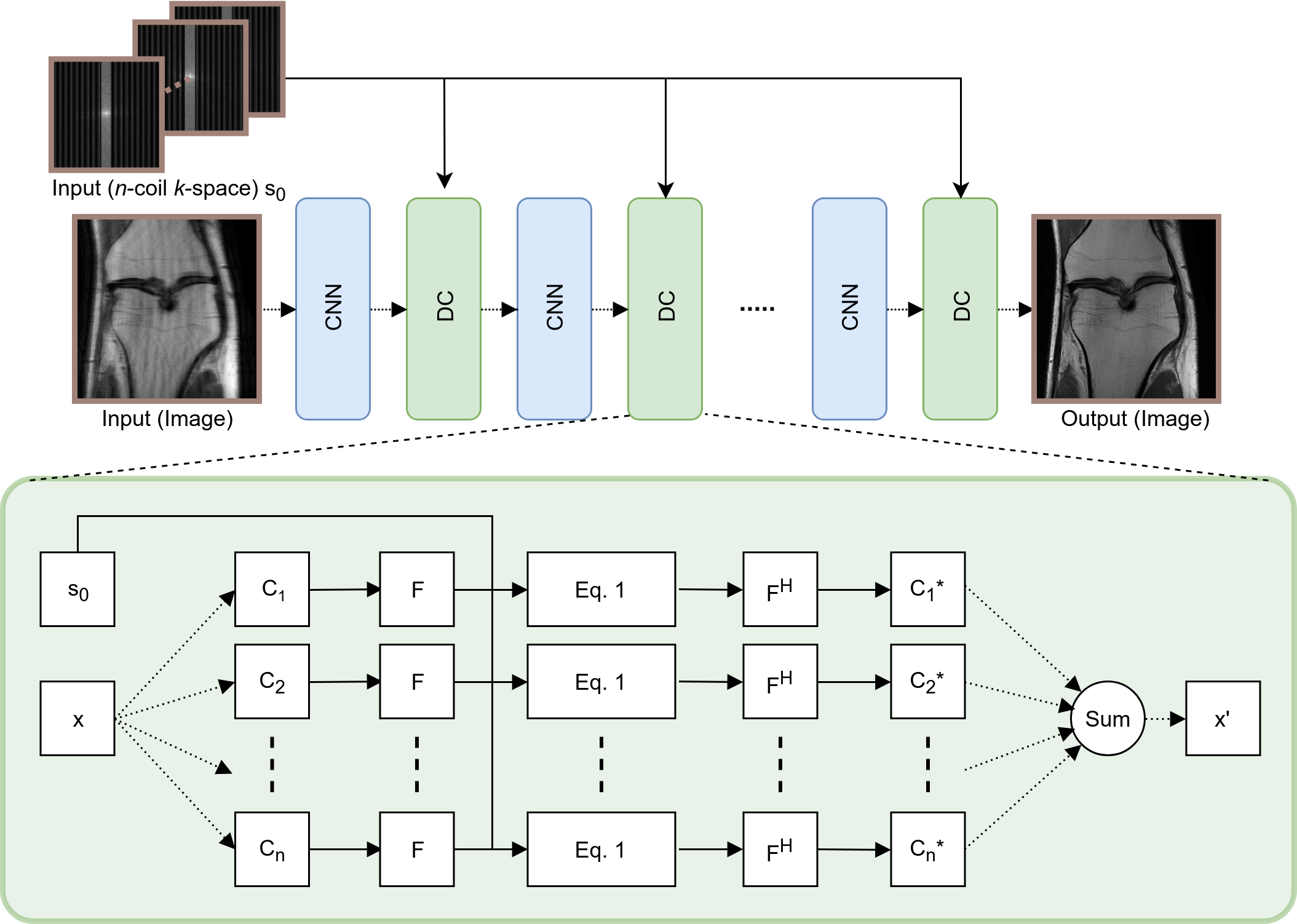}}
    \label{fig:architecture_dpocsense}\hfill
  \subfloat{
       \includegraphics[width=0.48\linewidth]{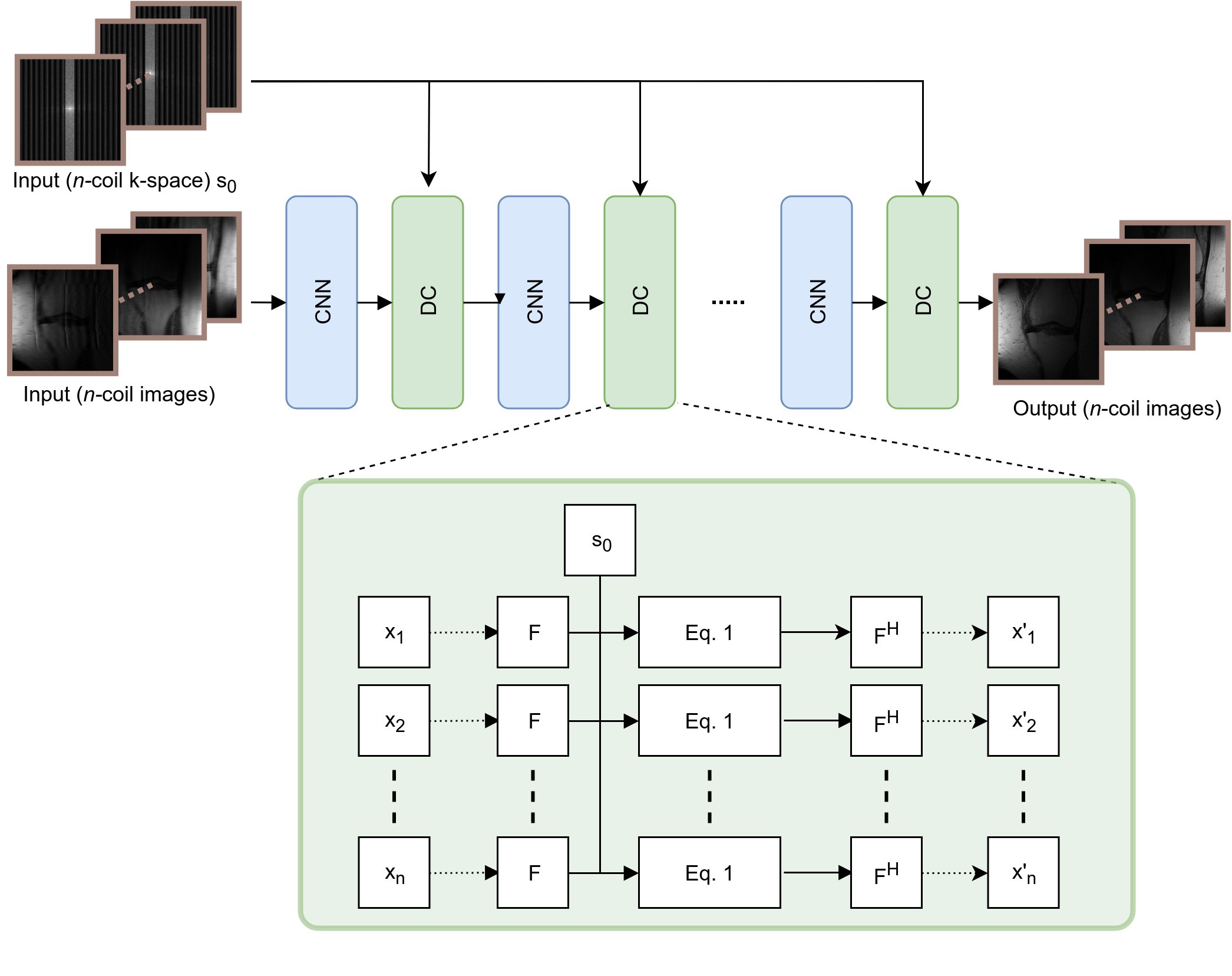}}
    \label{fig:architecture_dccnn}
  \caption{The proposed network architectures. (left) D-POCSENSE architecture. The input to the CNN is a single, sensitivity-weighted recombined image. At each iteration, the CNN updates an estimate of the combined image. The sub-network takes a single recombined image as an input and produces the denoised result as an output. The data consistency is performed by mapping the intermediate output to the raw $k$-space by applying encoding matrix. The updated image is recombined by the adjoint of the encoding matrix. (right) The proposed DC-CNN architecture. The network jointly reconstructs each coil data simultaneously. The data consistency operation is applied separately for each coil.}
  \label{fig:architectures} 
\end{figure}

The proposed networks are direct extensions of deep cascades of CNN (DC-CNN), where the denoising sub-networks and the data consistency layers are interleaved. However, for parallel imaging, the data consistency layer can be extended in two ways, yielding two network variants. In the first approach, sensitivity estimates are required, which can be computed using algorithms such as E-SPIRiT\cite{uecker2014espirit}. The input to the CNN is a single, sensitivity-weighted recombined image. At each iteration, the CNN updates an estimate of the combined image. For the data consistency layer, the forward operation is performed, then acquired samples are filled coil-wise as:
\begin{align}
\label{eq:dc}
s_\text{rec}^{(i)}(k) = \begin{cases}
\lambda s_\text{cnn}^{(i)}(k) + (1 - \lambda) s_0^{(i)} (k) & \text{ if } k \in \Omega \\
s_\text{cnn}^{(i)}(k) & \text{ if } k \not \in \Omega \\
\end{cases}
\end{align}
where $s_\text{cnn}^{(i)}$, $s_0^{(i)}$ are the $i$-th coil-weighted image for the intermediate CNN reconstruction in $k$-space and the original $k$-space data respectively. The result is mapped back to image domain via the adjoint of the encoding matrix. As the operation in the data consistency layer is analogous to the projection step from POCSENSE, the proposed network is termed D(eep)-POCSENSE. The balancing term $\lambda$ depends on the input noise level, however, this is made trainable as a network parameter. The network is trained using $\ell_2$ loss:
\begin{align}
\ell_2(\theta) = \sum_{(x_u, x) \in D} \| x - f_\text{cnn}(x_u; \theta) \|_2^2    
\end{align}
where $x_u$ and $x$ are the initial recombined image and ground truth respectively.

The second approach reconstructs the multiple coil data directly without performing the recombination and the coil images are stacked along the channel-axis and fed into each sub-network. For the data consistency layer, each coil image is Fourier transformed and \cref{eq:dc} is applied individually. As it does not require a sensitivity estimate, the proposed approach is calibration-less. The proposed network, DC-CNN, is trained with the following weighted-$\ell_2$ loss:
\begin{align}
\ell_2(\theta) = \sum_{(x_u, x) \in D} \sum_{i=1}^{n_\text{coil}} \| C_i^H \left(x_i - f_\text{cnn}(x_{u,1}, \dots, x_{u,n_\text{coil}}; \theta)_i \right) \|_2^2    
\end{align}
where the subscript indexes $i$-th coil data and $C_*^H$ is the sensitivity map. The proposed architectures are shown in \cref{fig:architectures}.

\section{Evaluation} 

We used the public knee dataset provided by Hammernik et al.\cite{hammernik2018learning}\footnote{Available at \url{mridata.org}.}. The dataset contains 100 patients, 20 subjects per acquisition protocol. For each approach, one network was trained to reconstruct all acquisition protocols simultaneously. We used 15 for training and 5 for testing per protocol. The proposed approach was compared with $\ell_1$-SPIRiT\cite{murphy2012fast} and Variational Network (VN)\cite{hammernik2018learning}. We used Cartesian undersampling with acceleration factor (AF)  $
\in \{4,6\}$, sampling 24 central region, which was also used as the calibration region for estimating the sensitivity maps. In this work, D-POCSENSE and DC-CNN were trained with $n_\text{coil}\!=\!15$,$n_d\!=\!5$,$n_c\!=\!10$\cite{schlemper2017deep} and convolution kernels with dilation factor 2. The network was trained using Adam with $lr\!=\!10^{-3}$ for 200 epoch with batch size 4. The default parameters were used for both $\ell_1$-SPIRiT and VN. We used PSNR and SSIM for the metric.

\begin{table}[t!]
\centering
\caption{The summary of quantitative results for AF=4 and AF=6. PD denotes proton tensity and FS denotes fat saturation.} 
\label{table:quantitative}
\begin{adjustbox}{width=\linewidth}
\begin{tabular}{clcccc}
\toprule
                     &            & \multicolumn{2}{c}{AF=4} & \multicolumn{2}{c}{AF=6} \\
\cmidrule(lr){3-4} \cmidrule(lr){5-6}
Modality             & Model      & PSNR        & SSIM      & PSNR        & SSIM      \\
\midrule
\multirowcell{4}{Axial $T_2$ FS}    & $\ell_1$-SPIRIT  & 33.55$\pm$12.62 & 0.87$\pm$0.26 & 26.61$\pm$12.83 & 0.80$\pm$0.30 \\
                     & D-POCSENSE & 36.03$\pm$10.26 & 0.90$\pm$0.24 & 31.83$\pm$8.79  & {\bf 0.87}$\pm${\bf 0.25} \\
                     & DC-CNN     & 35.45$\pm$8.63  & {\bf 0.91}$\pm${\bf 0.19} & 30.14$\pm$6.85  & {\bf 0.87}$\pm${\bf 0.23} \\
                     & VN         & {\bf 36.49}$\pm${\bf 10.41} & 0.90$\pm$0.23 & {\bf 32.39}$\pm${\bf 9.94}  & 0.86$\pm$0.29 \\
\midrule
\multirowcell{4}{Coronal PD}
                     & $\ell_1$-SPIRIT  & 36.96$\pm$1.28  & {\bf 0.98}$\pm${\bf 0.00} & 31.84$\pm$1.32  & {\bf 0.95}$\pm${\bf 0.01} \\
                     & D-POCSENSE & 36.94$\pm$1.24  & {\bf 0.98}$\pm${\bf 0.00} & 32.27$\pm$0.83  & {\bf 0.95}$\pm${\bf 0.00} \\
                     & DC-CNN     & 35.14$\pm$1.26  & 0.97$\pm$0.01 & 29.88$\pm$1.72  & 0.94$\pm$0.01 \\
                     & VN         & {\bf 37.07}$\pm$1.15  & {\bf 0.98}$\pm${\bf 0.00} & {\bf 33.17}$\pm${\bf 1.06}  & {\bf 0.95}$\pm${\bf 0.01} \\
\midrule
\multirowcell{4}{Coronal PDFS}
                     & $\ell_1$-SPIRIT  & 36.98$\pm$7.85  & 0.97$\pm$0.06 & 33.32$\pm$6.84  & 0.95$\pm$0.07 \\
                     & D-POCSENSE & 39.02$\pm$3.37  & {\bf 0.98}$\pm${\bf 0.02} & 34.91$\pm$2.7   & {\bf 0.97}$\pm${\bf 0.04} \\
                     & DC-CNN     & 38.37$\pm$3.23  & {\bf 0.98}$\pm${\bf 0.03} & 33.61$\pm$2.74  & 0.96$\pm$0.05 \\
                     & VN         & {\bf 39.39}$\pm${\bf 3.32}  & {\bf 0.98}$\pm${\bf 0.02} & {\bf 35.71}$\pm$2.80  & {\bf 0.97}$\pm${\bf 0.02} \\
\midrule
\multirowcell{4}{Sagittal PD}
                     & $\ell_1$-SPIRIT  & 36.76$\pm$0.47  & {\bf 0.98}$\pm${\bf 0.00} & 31.43$\pm$0.84  & 0.94$\pm$0.01 \\
                     & D-POCSENSE & 37.09$\pm$0.54  & {\bf 0.98}$\pm${\bf 0.00} & 31.94$\pm$0.45  & 0.94$\pm$0,00 \\
                     & DC-CNN     & 35.76$\pm$0.59  & 0.97$\pm$0.00 & 30.12$\pm$1.31  & 0.94$\pm$0.01 \\
                     & VN         & {\bf 37.47}$\pm${\bf 0.55} & {\bf 0.98}$\pm${\bf 0.00} & {\bf 32.86}$\pm${\bf 0.59}  & {\bf 0.95}$\pm${\bf 0.00} \\
\midrule
\multirowcell{4}{Sagittal $T_2$ FS}
                     & $\ell_1$-SPIRIT  & 37.71$\pm$1.55  & {\bf 0.98}$\pm${\bf 0.00} & 33.32$\pm$1.29  & {\bf 0.96}$\pm${\bf 0.01} \\
                     & D-POCSENSE & 37.96$\pm$1.08  & {\bf 0.98}$\pm${\bf 0.01} & 33.43$\pm$1.08  & {\bf 0.96}$\pm${\bf 0.01} \\
                     & DC-CNN     & 37.02$\pm$1.55  & {\bf 0.98}$\pm${\bf 0.01} & 27.76$\pm$3.04  & 0.95$\pm$0,01 \\
                     & VN         & {\bf 38.39}$\pm${\bf 1.21}  & {\bf 0.98}$\pm${\bf 0.01} & {\bf 34.32}$\pm${\bf 1.12}  & {\bf 0.96}$\pm${\bf 0.01} \\
\bottomrule
\end{tabular}
\end{adjustbox}
\end{table}

\begin{figure}[t!]
	\centering
  \subfloat{
       \includegraphics[width=0.48\linewidth]{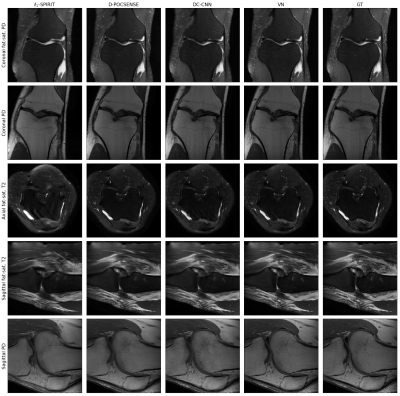}}
    \label{fig:qualitative_ac4}\hfill
  \subfloat{
       \includegraphics[width=0.48\linewidth]{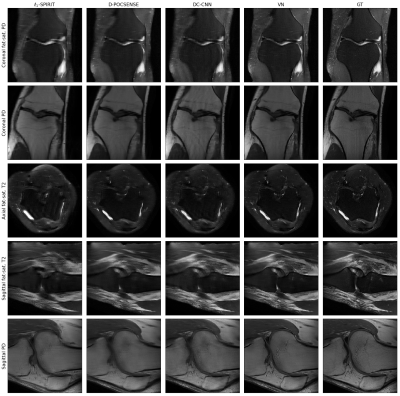}}
    \label{fig:qualitative_ac6}
  \caption{(left) The reconstruction results from each method for Cartesian undersampling (left) with acceleration factor 4 and (right) acceleration factor 6.}
  \label{fig:qualitative} 
\end{figure}

\section{Results}

Quantitative results are summarised in \cref{table:quantitative} for each acquisition protocol. The proposed methods both outperformed the compressed sensing approach on average. D-POCSENSE achieved the performance close to VN for AF=4, whereas DC-CNN was slightly worse. All methods provided similar SSIM. For AF=6, VN achieved the highest PSNR. The sample reconstructions are shown in \cref{fig:qualitative} for AF=4 and AF=6 respectively. For Axial $T_2$ image, D-POCSENSE gave the most homogeneous image, whereas DC-CNN and VN often failed to remove aliasing. For AF=4, all methods generated sharp images. For AF=6, DC-CNN performed worse that D-POCSENSE and VN and the residual aliasing is prominent.

\section{Discussion and Conclusion}

In this work, we proposed simple extensions to DC-CNN for parallel imaging. When comparing the two approaches so far explored, D-POCSENSE outperformed DC-CNN overall, which leads to the conclusion that incorporating the sensitivity estimate is advantageous. We speculate that this is because it allows intermediate sub-networks to directly operate in the output space as well as directly optimising the loss with respect to the final output. Nevertheless, DC-CNN achieved the highest SSIM in some regimes, which shows that a novel way of combining the raw data could lead to improved algorithms. The proposed methods achieved comparable performance to state-of-the-art algorithms, however, we note that the variational network produced the best result overall.

\section{Note}

We observed that training D-POCSENSE and DC-CNN networks longer can further remove the residual aliasing present in the reconstruction to eventually reach similar performances. The presented work is now extended to \emph{variable-splitting network} \cite{duan2019vs}.

\section{Acknowledgements}

Jo Schlemper is partially funded by EPSRC Grant (EP/P001009/1).

\bibliographystyle{splncs03}
\bibliography{ref}

\end{document}